\documentclass[prb,twocolumn,showpacs,preprintnumbers,amsmath,amssymb,superscriptaddress,floatfix]{revtex4-2}

\usepackage[makeroom]{cancel}

\usepackage{amsmath}    
\usepackage{amssymb,environ}
\usepackage{graphicx}   
\usepackage{verbatim}   
\usepackage{color}      
\usepackage{hyperref}   
\usepackage{breakurl}
\usepackage[normalem]{ulem}
\usepackage{natbib}
\usepackage{enumitem}
\usepackage{multirow}

\usepackage{epsfig}
\usepackage{textcomp}
\usepackage{wasysym}
\usepackage{array}
\usepackage{color}

\begin{document}

\newcommand{\ev}[0]{\mathbf{e}}
\newcommand{\cv}[0]{\mathbf{c}}
\newcommand{\fv}[0]{\mathbf{f}}
\newcommand{\Rv}[0]{\mathbf{R}}
\newcommand{\Tr}[0]{\mathrm{Tr}}
\newcommand{\ud}[0]{\uparrow\downarrow}
\newcommand{\Uv}[0]{\mathbf{U}}
\newcommand{\Iv}[0]{\mathbf{I}}
\newcommand{\Hv}[0]{\mathbf{H}}

\setlength{\jot}{2mm}

\newcommand{\jav}[1]{{\color{red}#1}}

\title{Exchange interaction between two quantum dots coupled through a superconducting island}

\author{\'Ad\'am B\'acsi}
\email{bacsi.adam@sze.hu}
\affiliation{Jo\v zef Stefan Institute, Jamova 39, Ljubljana SI-1000, Slovenia}
\affiliation{Department of Mathematics and Computational Sciences, Sz\'echenyi Istv\'an University, 9026 Gy\H or, Hungary}

\author{Luka Pave\v si\' c}
\affiliation{Jo\v zef Stefan Institute, Jamova 39, Ljubljana SI-1000, Slovenia}
\affiliation{Faculty of Mathematics and Physics, University of Ljubljana, Jadranska 19, SI-1000 Ljubljana, Slovenia}

\author{Rok \v Zitko}
\affiliation{Jo\v zef Stefan Institute, Jamova 39, Ljubljana SI-1000, Slovenia}
\affiliation{Faculty of Mathematics and Physics, University of Ljubljana, Jadranska 19, SI-1000 Ljubljana, Slovenia}

\date{\today}

\begin{abstract}
We present a theoretical study of a system consisting of a superconducting island and two quantum dots, a possible platform for building qubits and Cooper pair splitters, in the regime where each dot hosts a single electron and, hence, carries a magnetic moment. We focus on the case where the dots are coupled to overlapping superconductor states and we study whether the spins are ferromagnetically or antiferromagnetically aligned. We show that if the total number of particles is even, the spins align antiferromagnetically in the flatband limit, i.e., when the bandwidth of the superconductor is negligibly small, but ferromagnetically if the bandwidth is finite and above some value. If the total number of particles is odd, the alignment is ferromagnetic independently from the bandwidth. This implies that the results of the flatband limit are applicable only within restricted parameter regime for realistic superconducting qubit systems and that some care is required in applying simplified models to devices such as Cooper pair splitters.
\end{abstract}

\maketitle

\section{Introduction}

Superconducting devices are attracting strong interest due to their applications in quantum computing \cite{yu2002,koch2007,Devoret2013,wendin2017,arute2019,annualreview2020,clerk2020,Aguado2020,blais2021,siddiqi2021,hendrickx2021}. Building upon the ideas of semiconductor spin qubits \cite{taylor2005, burkard2021}, the combination of spin carrying quantum dots and superconducting islands is also a possible qubit realization \cite{karrasch2011, padurariu2010, rok2022, rokluka2022, filip} which remains to be explored more in depth.Such devices are particularly promising due to the presence of tunable long-lived subgap states. They could find use in a wide range of applications, such as Cooper-pair splitters \cite{kurtossy2021}, for linking spatially separated superconducting qubits \cite{petersson2012,mi2017,borjans2019}, or building Andreev molecules \cite{pillet2019}.

Quantum dots (QDs) can be designed to possess a single relevant electron state well separated from other levels \cite{zwanenburg2013}. In this case, the QD may be modeled as a single Anderson impurity with two characteristic energy scales:  the energy of the QD level, $\varepsilon$, and the on-site interaction energy, $U$. If $U$ is large and $\varepsilon$ is deep below the Fermi level, the QD hosts a single electron and carries a spin just like magnetic impurities.

Two  magnetic impurities in the same host material can be coupled in different ways. In a metallic host, they couple via the Ruderman-Kittel-Kasuya-Yosida (RKKY) interaction which, depending on the inter-impurity distance, may result in a ferromagnetic (FM) or antiferromagnetic (AFM) spin alignment. For very short distances, when the distance is on the order of the lattice constant, the atomistic details play a crucial role and the coupling is rather described as superexchange with the coupling sign determined by the electron filling of the intermediary atoms. Superexchange is also the essential mechanism governing the interaction between non-neighboring QDs of a chain \cite{kandel2021apl}.

In the case of superconducting mediator, the RKKY interaction is shifted toward AFM alignment \cite{glazmanprl2014} due to additional processes involving the subgap states. The effect of superconductivity on superexchange, however, has not been studied much so far.

\begin{figure}[ht]
\centering
\includegraphics[width=6cm]{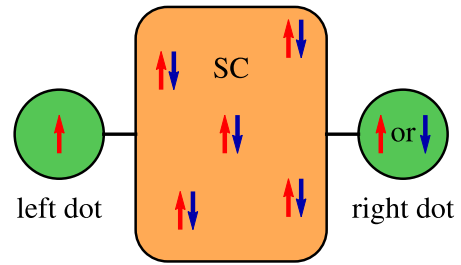}
\caption{Schematic of a superconducting island bridging two quantum dots. Each dot carries a spin.}
\label{fig:qdscqd}
\end{figure}

In this paper, we study the system with two QDs coupled to a superconducting island (SI) as sketched in Fig.~\ref{fig:qdscqd}. 
Our goal is to elucidate the sign and the strength of the inter-dot exchange coupling in the case where the two dots are coupled to the same superconductor state: this corresponds to the limit of small separation between the dots.

\section{Model of the superconductor}

Conventional superconductors are successfully described by the BCS mean-field theory \cite{bcs}. An alternative approach to superconductivity consists of solving the pairing model without performing the mean-field decoupling in the canonical ensemble. The simplified version of the pairing Hamiltonian is also known as the Richardson or the picket-fence model \cite{vonDelft2001,dean2003,Tempere2019}. In the original work of Richardson and Sherman \cite{Richardson1964}, several energy levels $\epsilon_i$ are taken into account which, in the context of solid state physics, can be thought of as the states in the vicinity of the Fermi energy that participate in the attractive pairing interaction. The pairing interaction consists of quartic terms [see Eq.~\eqref{eq:rpairing}] leading to the formation of Cooper pairs.

The Richardson model is exactly solvable for an isolated superconductor (SC). However, if the SC is coupled to QDs, exact analytical solutions are available only in the flatband limit. In this limit, the energy levels are all degenerate, $\epsilon_i \equiv \text{const}$. Despite of this simplification, the flatband limit was proven to provide a proper description for systems of one QD coupled to the SI \cite{rok2022}. In this paper, we extend these studies to the case of two QDs coupled to the island.

Let us consider a system with $N$ fermionic single-particle levels and we write the pairing Hamiltonian of the Richardson model as
\begin{gather}
H_{SC}=-\frac{g}{N}\sum_{i,j}^N c_{i\uparrow}^+ c_{i\downarrow}^+ c_{j\downarrow} c_{j\uparrow},
\label{eq:rpairing}
\end{gather}
where $c_{i\sigma}$ is the fermionic annihilation operator corresponding to the energy level $i$ and spin $\sigma$. In contrast to the BCS Hamiltonian, the number of fermions is conserved. For $N$ superconducting levels and $M$ Cooper pairs, the ground state is given by $|\Psi_{M}^N\rangle = \sum_{conf}|\mathrm{conf}\rangle/\sqrt{\mathcal{N}_{M}^N}$ which is the completely symmetric combination of all possible configurations of $M$ Cooper pairs occupying the $N$ levels. The number of configurations is $\mathcal{N}_M^N=\binom{N}{M}$. The ground state (GS) energy is  expressed as\cite{Tempere2019,rok2022}
\begin{gather}
E_{0} = -\frac{g}{N}M(N+1-M).
\end{gather}

A great advantage of the flatband limit is that the model is invariant under any unitary transformations as long as the opposite spin levels are transformed with complex conjugate matrices. To be more precise, for any unitary matrix $U_{ij}$, the Hamiltonian is invariant under 
\begin{equation}
    \begin{split}
f_{j\uparrow} &= \sum_{i}U_{ji}c_{i\uparrow}, \\
f_{j\downarrow} &= \sum_i U_{ji}^* c_{i\downarrow}.
    \end{split}
\end{equation}
The physical significance of the complex conjugation is that it maintains the time-reversal invariance of the Hamiltonian.

At this point, we have complete freedom how to choose the unitary transformation. We will use this freedom to transform into a basis in which the hybridization between the dots and the superconductor has a convenient form. In general, the hybridization operators are defined as
\begin{gather}
H_{hyb} = H_{hyb,L} + H_{hyb,R},
\end{gather}
where
\begin{gather}
H_{hyb,L} = v_L \sum_{i} \left(\gamma_{Li} d_{L\uparrow}^+ c_{i\uparrow} + \gamma_{Li}^* d_{L\downarrow}^+ c_{i\downarrow} + h.c. \right)
\label{eq:lhyb}
\end{gather}
describes the coupling between the left dot and the superconductor. $d_{L\sigma}$ is the fermionic annihilation operator corresponding to the left dot level. The strength of the coupling is determined by $v_L$ while the coefficients $\gamma_{Li}$ describe the distribution of the couplings obeying $\sum_{i}|\gamma_{Li}|^2 = 1$ (i.e., they define the wave function of the superconductor state that hybridizes with the QD level). Note that for different spin orientations, we use complex conjugate coupling constants as a consequence of the time-reversal symmetry.
The coupling between the right dot and the superconductor is given by the same formula as \eqref{eq:lhyb}, but with $v_R$ and $\gamma_{Ri}$. The coefficients are also normalized as $\sum_i|\gamma_{Ri}|^2=1$. 

The overlap between the distribution of $\gamma_{Li}$ and $\gamma_{Ri}$ is given by \cite{filip}
\begin{equation}
    \alpha=\sum_i \gamma_{Li}^*\gamma_{Ri}
\end{equation}
whose absolute value ranges from 0 to 1. Within the framework of the Richardson model, $\alpha$ mimics the \textit{distance} between the dots. Namely, $\alpha=0$ means that there is no overlap between the distributions and the dots are \textit{far} from each other. On the other hand, $|\alpha| = 1$ implies that the dots are coupled to the same superconducting state and are, hence, \textit{close} to each other. In the present paper, we focus on the latter limit, i.e., when the distributions $\gamma_{Li}$ and $\gamma_{Ri}$ are basically the same apart from an overall phase factor. This limit is valid as long as the Yu-Shiba-Rusinov states of the dots have a large overlap. Recent experiments show that the spatial extension of these states can reach even the range of 50 nm \cite{scherubl2020}.
We also note that $\alpha$ fully characterizes the relation between the coefficients $\gamma_{Li}$ and $\gamma_{Ri}$ in the flatband limit, while for finite bandwidth their detailed values (energy dependence) also matter.

As shown in the Appendix~\ref{appA}, there exists a unitary transformation such that $f_{0\uparrow} = \sum_i \gamma_{L/R i}c_{i\uparrow}$ is one of the new basis states which forms an orthonormal system together with the other new basis states. The state corresponding to $f_{0\sigma}$ will be referred to as the distinguished state of the superconductor.
The unitary transformation leaves the pairing Hamiltonian invariant in form so that 
\begin{gather}
H_{SC} = -\frac{g}{N}\sum_{ij} f_{i\uparrow}^+f_{i\downarrow}^+ f_{j\downarrow} f_{j\uparrow}
\label{eq:hamf}
\end{gather}
due to the flatband limit. In the presence of a non-zero bandwidth, the superconductor levels would have non-uniform $\epsilon_i$ which, under the unitary transformation, would generate inter-level hopping terms.

With the distinguished level $f_{0\sigma}$, let us define the many-body superconductor states
\begin{gather}
|\Psi_0\rangle = |0,\Psi_M^{N-1}\rangle \quad\mbox{and} \quad  |\Psi_2\rangle = |\ud,\Psi_{M-1}^{N-1}\rangle,
\label{eq:1outstates}
\end{gather}
where $\Psi_0$ ($\Psi_2$) describes the state where the $f_0$-level is empty (occupied by a Cooper pair) and the remaining $N-1$ superconductor levels are filled with $M$ (with $M-1$) Cooper pairs in a completely symmetric fashion.
The pairing Hamiltonian is represented on the basis of these states as
\begin{gather}
\Hv = E_{0} \Iv + \Hv_{SC}, \nonumber \\
\Hv_{SC} = \left[\begin{array}{cc} g\nu & \beta \\ \beta & g(1-\nu) \end{array}\right],
\label{eq:scmin}
\end{gather}
where 
\begin{equation}
\nu = M/N
\end{equation}
is the filling factor and 
\begin{equation}
\beta = -g\sqrt{\nu(1-\nu)}.
\label{eq:beta}
\end{equation}
The eigenvalues of the matrix are $0$ and $g$ corresponding to the GS and the excited state (ES), respectively. This indicates that the gap of the model is related to the pairing energy as $2\Delta = g$. Of course, there are many other states with the excitation energy of $g$, but this is the only one mixing with the GS in the presence of hybridization. The minimal model of the superconductor, Eq.~\eqref{eq:scmin}, is sufficient to study the coupling of two dots with the overlap parameter $\alpha = 1$ in the flatband limit. Note that if $\alpha\neq 1$, there would exist two distinguished levels (see Appendix~\ref{appA}) and instead of Eq.~\eqref{eq:1outstates} we would need to define a more complex set of many-body basis states.

\section{Two quantum dots coupled to the superconductor}

We study the effect of hybridization on the electronic structure in the case of $\alpha = 1$, when the two dots are coupled to the same superconductor level. Furthermore, it is assumed that the system has parity symmetry, i.e., $v_L = v_R = v$, $U_L=U_R=U$ and $\varepsilon_L = \varepsilon_R = \varepsilon$.

The total Hamiltonian of the system is expressed as
\begin{gather}
H=H_{SC} + H_{QD} + H_{hyb},
\end{gather}
where $H_{SC}$ is given by Eq.~\eqref{eq:hamf},
\begin{gather}
H_{QD} = \varepsilon \sum_{\sigma}\left(\hat{n}_{L\sigma} + \hat{n}_{R\sigma}\right) + U\left(\hat{n}_{L\uparrow}\hat{n}_{L\downarrow} + \hat{n}_{R\uparrow}\hat{n}_{R\downarrow}\right),
\end{gather}
and the coupling between the dots and the superconductor is determined by
\begin{gather}
H_{hyb} = \sum_{\sigma}\left[v\left(d_{L\sigma}^+ + d_{R\sigma}^+\right) f_{0\sigma} + h.c.\right],
\end{gather}
where $d_{L\sigma}$ ($d_{R\sigma}$) is the annihilation operator of the left (right) quantum dot.

The total Hamiltonian conserves both the number of particles and the total spin. In the followings, we determine the lowest-energy states in different charge and spin sectors by using analytical and numerical methods. 

In the analytical calculation, we apply the approximation that the on-site interaction $U$ is infinitely large and, hence, none of the dots can be doubly occupied. Furthermore, we take the thermodynamic limit of macroscopically large number of states in the SI. 

Numerical calculations are performed with Density Matrix Renormalization Group (DMRG) technique \cite{white1992,schollwock2011} in which $U$ has a finite value and the system is finite, $N=80$. For this purpose, we extended the approach described in Refs.~\onlinecite{coulomb1,coulomb2} and implemented a matrix product operator expression of the Hamiltonian for the QD-SI-QD system; see Appendix~\ref{MPO} for details.

We start the presentation of the analytical model by introducing four different sectors which are not coupled to each other because of the conservation of particle number and spin. For even number of particles, the total spin is either $0$ (singlet) or $1$ (triplet) since the superconductor is an overall singlet, but the electrons hosted by the dots can have arbitrary spin orientations. For odd number of particles, the total spin of the system is always $1/2$ (doublet), but based on the parity symmetry of states, a symmetric and an anti-symmetric sector can be identified.

\subsection{Analytical results for even number of particles}

In the sector of {\textit{even}} number of particles and $S=0$, the relevant states are given by
\begin{equation}
\begin{split}
|\phi_0^S\rangle &= \frac{1}{\sqrt{2}}\left( |\uparrow,\downarrow\rangle - |\downarrow,\uparrow\rangle \right) \otimes |0,\Psi_{M}^{N-1}\rangle, \\
|\phi_2^S\rangle &= \frac{1}{\sqrt{2}}\left( |\uparrow,\downarrow\rangle - |\downarrow,\uparrow\rangle \right) \otimes |\ud,\Psi_{M-1}^{N-1}\rangle, \\
|\phi_{0X}^S\rangle &= \frac{1}{2}\Big((|\uparrow,0\rangle + |0,\uparrow\rangle) \otimes |\downarrow,\Psi_{M}^{N-1}\rangle  \\ & \quad - (|0,\downarrow\rangle + |\downarrow,0\rangle) \otimes |\uparrow,\Psi_{M}^{N-1}\rangle \Big), \\
|\phi_{0D}^S\rangle &= |0,0\rangle \otimes |\ud,\Psi_{M}^{N-1}\rangle, \\
|\phi_{-2D}^S\rangle &= |0,0\rangle  \otimes |0,\Psi_{M+1}^{N-1}\rangle.
\label{eq:singletstates}
\end{split}
\end{equation}
The first part always encodes the state of the dots while the second part after $\otimes$ describes the state of the superconductor. 
It is also worth to note that all states in this sector are symmetric under the spatial reflection $L\leftrightarrow R$.
On this basis, the Hamiltonian is represented by the block matrix
\begin{gather}
\mathbf{H}_{S} = E_{0}\Iv + \left[\begin{array}{c|c|c} 
2\varepsilon + \Hv_{SC} & \begin{array}{c} \sqrt{2}v \\ 0 \end{array} & \text{\large 0} \\
\hline
\begin{array}{cc} \sqrt{2} v & 0 \end{array} & \varepsilon + g\nu & \begin{array}{cc} 2v & 0 \end{array} \\
\hline
\text{\large 0} & \begin{array}{c} 2v \\ 0 \end{array} & \Hv_{SC} + g(2\nu -1)\end{array}\right],
\label{eq:hams}
\end{gather}
where $\Hv_{SC}$ is the $2\times 2$ matrix describing the internal structure of the superconductor as defined in Eq. \eqref{eq:scmin}. We note that for finite system size, the matrix elements of the Hamiltonian would include further terms proportional to $1/N$ which, however, are neglected in the thermodynamic limit that we take here.

\begin{figure*}[htbp]
\centering
\includegraphics[width=16cm]{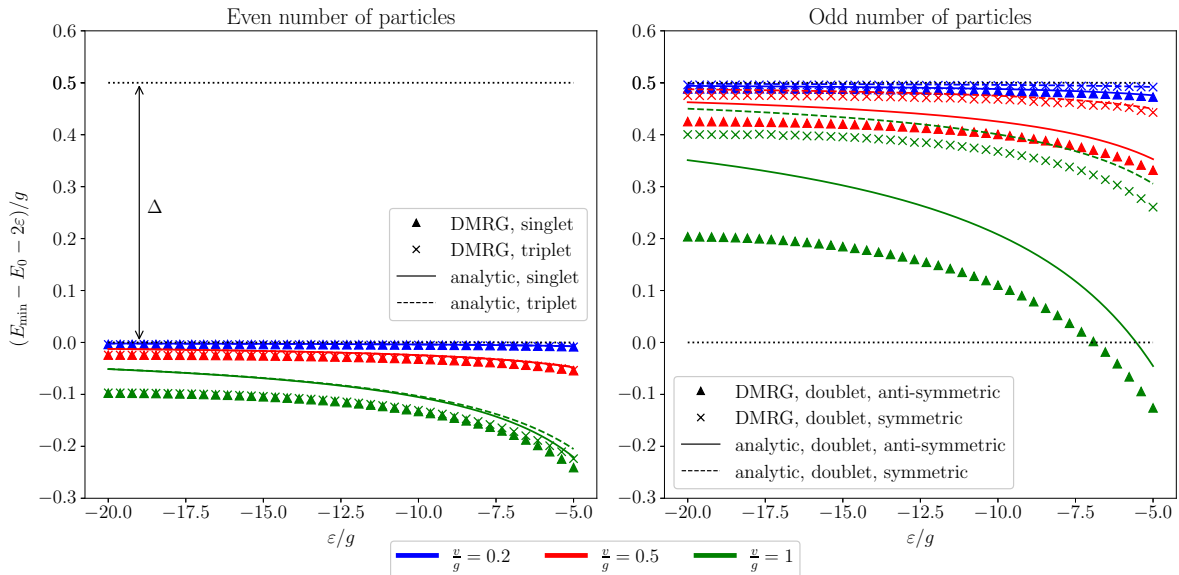}
\caption{Comparison of DMRG (denoted by symbols $\times$ and $\blacktriangle$) and analytical results (solid and dashed lines). Different colors correspond to different hybridization strengths, $v/g=0.2$, $v/g=0.5$ and $v/g=1$.}
\label{fig:flatband}
\end{figure*}

In the S=1 sector, we consider the $S_z=1$ states, other $S_z$ subspaces being equivalent.
The basis vectors are given as
\begin{equation}
\begin{split}
|\phi_0^T\rangle &= |\uparrow,\uparrow\rangle \otimes |0,\Psi_{M}^{N-1}\rangle, \\
|\phi_2^T\rangle &= |\uparrow,\uparrow\rangle \otimes |\ud,\Psi_{M-1}^{N-1}\rangle, \\
|\phi_{0X}^T\rangle &= \frac{1}{\sqrt{2}}\left(|\uparrow,0\rangle - |0,\uparrow\rangle\right) \otimes |\uparrow,\Psi_{M}^{N-1}\rangle,
\label{eq:tripletstates}
\end{split}
\end{equation}
which are all parity anti-symmetric states.
On this subspace, the Hamiltonian is represented by
\begin{gather}
\mathbf{H}_T=E_{0}\Iv + \left[\begin{array}{c|c} 
2\varepsilon + \Hv_{SC} & \begin{array}{c} \sqrt{2}v \\ 0 \end{array} \\
\hline
\begin{array}{cc} \sqrt{2} v & 0 \end{array} & \varepsilon + g \nu
\end{array}\right]\,.
\label{eq:hamt}
\end{gather}
To compare the two sectors with an even number of particles, we compute the ground state of the Hamiltonian in each sector as a function of $\varepsilon$. The results shown in Fig.~\ref{fig:flatband} indicate that the $S=0$ sector always has lower minimal energy. In fact, one can prove using perturbation theory that for $v\ll |\varepsilon|$ the difference of minimal energies 
\begin{equation}
E_T-E_S = \frac{2v^4}{|\varepsilon|^3}
\end{equation}
is always positive. The scaling of $v^4$ implies that the energy difference stems from a fourth order process which is characteristic to both superexchange and to RKKY interaction. The fourth order processes involve depopulation of the dots in the singlet sector through the states of $\psi_{0D}^S$ and $\phi_{-2D}^S$. However, in the triplet sector, the dots cannot be completely depopulated due to Pauli principle and because single particles cannot move from the distinguished level to other SC levels in the flatband limit. 

\subsection{Analytical results for odd number of particles}
If the total number of particles is {\textit{odd}}, the system has an overall spin of $S=1/2$ describing a {\textit{doublet}} state. 
Here, we take $S_z=1/2$. 
This sector can be split into a symmetric and an anti-symmetric subsector based on parity symmetry. 

In the {\textit{symmetric}} subsector, the relevant states are given by
\begin{equation}
\begin{split}
|\phi_1^{Dsym}\rangle &= \frac{1}{\sqrt{2}}\left( |\uparrow,\downarrow\rangle - |\downarrow,\uparrow\rangle\right) \otimes |\uparrow,\Psi_{M-1}^{N-1}\rangle , \\
|\phi_0^{Dsym}\rangle &= \frac{1}{\sqrt{2}}\left( |\uparrow,0\rangle + |0,\uparrow\rangle\right) \otimes |0,\Psi_{M}^{N-1}\rangle , \\
|\phi_2^{Dsym}\rangle &= \frac{1}{\sqrt{2}}\left( |\uparrow,0\rangle + |0,\uparrow\rangle\right) \otimes |\ud,\Psi_{M-1}^{N-1}\rangle , \\
|\phi_{-1}^{Dsym}\rangle & = |0,0\rangle \otimes |\uparrow,\Psi_{M}^{N-1}\rangle ,
\end{split}
\end{equation}
and the Hamiltonian is represented by
\begin{gather}
\Hv^{sym}_D = E_{0}\Iv + \left[\begin{array}{c|c|c} 2\varepsilon + g(1-\nu) & \begin{array}{cc} 0 & -v \end{array} & 0 \\ 
\hline
\begin{array}{c} 0 \\ -v \end{array} & \varepsilon + \Hv_{SC} & \begin{array}{c} \sqrt{2}v \\ 0 \end{array} \\
\hline
0 & \begin{array}{cc} \sqrt{2}v & 0 \end{array} & g\nu \end{array}\right]\,.
\label{eq:hamdsym}
\end{gather}

In the anti-symmetric subsector, the relevant states are
\begin{equation}
\begin{split}
|\phi_{1}^{Dasym}\rangle &= \frac{1}{\sqrt{6}}\left( 2|\uparrow,\uparrow\rangle\otimes |\downarrow,\Psi_{M-1}^{N-1}\rangle \right. - \nonumber \\ & - \left.\left(|\uparrow,\downarrow\rangle + |\downarrow,\uparrow\rangle\right) \otimes |\uparrow,\Psi_{M-1}^{N-1}\rangle \right), \\
|\phi_0^{Dasym}\rangle &= \frac{1}{\sqrt{2}}\left( |\uparrow,0\rangle - |0,\uparrow\rangle\right) \otimes |0,\Psi_{M}^{N-1}\rangle , \\
|\phi_2^{Dasym}\rangle &= \frac{1}{\sqrt{2}}\left( |\uparrow,0\rangle - |0,\uparrow\rangle\right) \otimes |\ud,\Psi_{M-1}^{N-1}\rangle ,
\end{split}
\end{equation}
and the Hamiltonian is given by
\begin{gather}
\Hv^{asym}_D = E_{0}\Iv + \left[\begin{array}{c|c} 
2\varepsilon + g(1-\nu) & \begin{array}{cc} 0 & \sqrt{3}v \end{array} \\
\hline
\begin{array}{c} 0 \\ \sqrt{3}v \end{array} & \varepsilon + \Hv_{SC}
 \end{array}\right]\,.
\label{eq:hamdasym}
\end{gather}
The minimal energies are numerically obtained and shown in Fig.~\ref{fig:flatband}. The results show that the anti-symmetric sector has lower minimal energies which is further confirmed by the perturbative expansion for $|\varepsilon|\gg v$ leading to 
\begin{equation}
    E_{D}^{sym}- E_{D}^{asym} = \frac{2v^2}{|\varepsilon|}.
\end{equation}
Note that the anti-symmetric sector, particularly the state $|\phi_1\rangle$, features FM alignment of the spins sitting on the dots even though the total system is a spin-1/2 state.
The FM coupling between the spins differs essentially from what has been found in the case of even number of particles, where the spins show AFM alignment in the singlet sector.

\subsection{Discussion}

The difference in behavior for odd and even electron number can be understood by recalling the properties of superexchange \cite{anderson1950,goodenough1955}. If the superexchange between two magnetic moments is realized through a single orbital level, it can be shown that the nature of coupling depends on the occupancy of the mediating orbital, see Refs.~\cite{anderson1950,anderson1959} and  the Appendix~\ref{appC}. If it is singly occupied, the magnetic moments are preferentially FM aligned, while for empty or doubly occupied mid-state, the coupling is AFM. For an odd total number of particles, the energetically most favourable configuration is that the extra particle occupies the distinguished level of the superconductor ensuring single occupancy in the orbital that bridges both quantum dot levels.

For even number of particles, however, all electrons of the superconductor tend to form Cooper pairs and, hence, the distinguished level is more likely to be empty or doubly occupied. We note that the probability of having a single particle on the distinguished level is zero only in the strict flatband limit. For finite bandwidth, single particle hopping takes place among the SC levels which indicates that the probability of single occupancy is not zero. This effect is further analyzed in Sec. \ref{sec:fbw}.

\subsection{Comparison with DMRG results}
We compare the analytical results with the results of a DMRG calculation carried out with $N=80$ superconducting levels and $n=82$, 83 particles. The numerical simulation has been performed with a uniform distribution of couplings, $\gamma_{Li}=\gamma_{Ri} =1/\sqrt{N}$ which corresponds to $\alpha = 1$ as explained after Eq.~\ref{eq:lhyb}. The on-site interaction has been set to $U/g=40$.

The comparison in Fig.~\ref{fig:flatband} confirms that the analytical approach produces correct results as regards the spin configuration. Namely, numerical simulations also indicate AFM coupling for even and FM coupling for odd number of particles. However, some quantitative differences are observed. These are the consequence of finite $U$ and of the finite system size in DMRG.
 The dot energy level has been swept between $-20g$ and $-5g$ to cover the regime where the probability of single occupancy of the QDs is high. We do not study $\varepsilon<-20g$ because charge fluctuations through doubly occupied states, which are not included in our analytical model, would play a more important role. The fingerprints of these charge fluctuations are captured by the DMRG close to $\varepsilon = -U/2 = -20g$ but these have no effect on the spin configuration.




\section{Effects of finite bandwidth}
\label{sec:fbw}
The flatband limit has been shown to represent an adequate model for the situation of a single QD coupled to the SI \cite{rok2022}. There, the effects of finite bandwidth lead to some quantitative modifications but no change in the general structure has been observed. 

We now investigate whether the nature of spin couplings of the double dot system uncovered in the flatband limit is preserved for finite bandwidth. Since the Richardson model with finite bandwidth and attached QDs cannot be handled analytically, we mostly rely on DMRG calculations (below we will also introduce a minimal analytical model which qualitatively describes the main new feature). For simplicity, we will consider the case of a uniform distribution of energy levels $\epsilon_i$ in the SI spanning the interval $[-W:W]$. The numerical simulations are performed with several hybridization strengths, $v/g= 0.2$, $v/g = 0.5$ and $v/g = 1$. The QD parameters are $U/g=40$ and $\varepsilon/g = -10$. 

We find that for an odd number of particles the finite bandwidth does not change the spin configuration in the ground state. In the studied regime, the spins are always FM aligned.

\begin{figure}[ht]
\centering
\includegraphics[width=8cm]{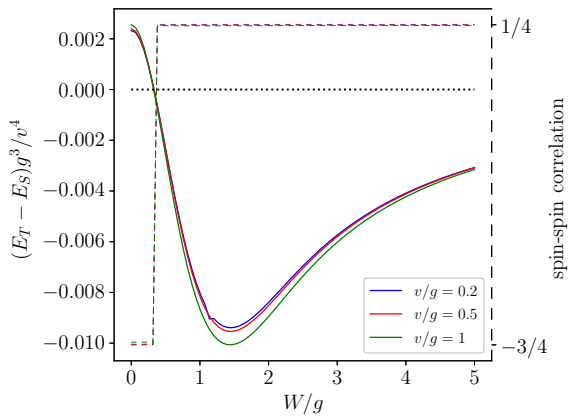}
\caption{Effects of finite bandwidth for even number of particles at different values of hybridization strength, DMRG results. The energy difference $E_T-E_S$ (solid lines, left vertical axis) changes sign  with increasing bandwidth indicating that the ground state becomes a triplet state. This feature is further verified by calculating the spin-spin correlation between the dots (dashed lines, right vertical axis).}
\label{fig:finbw}
\end{figure}

However, for the even number sector, we do find a difference: whereas the flatband limit exhibits AFM alignment, we observe FM alignment at large enough values of the bandwidth. Fig.~\ref{fig:finbw} shows that above some limit value of the bandwidth, which is in the range of $g$, the triplet sector has lower energy.

We now present an analytical model which recovers this feature. The finite bandwidth is the consequence of the kinetic energy of the particles in the superconductor. In the original basis determined by the operators $c_i$, the kinetic energy is represented by a diagonal matrix with the entries of $\epsilon_i$. After applying the unitary transformation $c_i\rightarrow f_j$ transforming the hybridization into a more convenient form, the kinetic energy term becomes non-diagonal. In principle all-to-all couplings could appear but we only consider those in which the distinguished state $f_0$ participates.
Therefore, the kinetic energy term is written as
\begin{gather}
H_{kin} = \frac{t}{\sqrt{N}}\sum_{i\neq 0,\sigma}\left(f_{i\sigma}^+ f_{0\sigma} + h.c.\right)\,.
\end{gather}

This hopping process breaks Cooper pairs inside the superconductor. Therefore, the basis set of $|\Psi_0\rangle$ and $|\Psi_2\rangle$ as defined in Eq.~\eqref{eq:1outstates} must be complemented with the Cooper-pair-broken state
\begin{gather}
|\Psi_1\rangle = \frac{1}{\sqrt{2}}\Big( |\uparrow,\Psi_{M-1,+\downarrow}^{N-1}\rangle - |\downarrow,\Psi_{M-1,+\uparrow}^{N-1}\rangle\Big),
\end{gather}
where
\begin{gather}
|\uparrow,\Psi_{M-1,+\downarrow}^{N-1}\rangle = \frac{1}{\sqrt{N-1}}\sum_{i=1}^{N-1}f_{0\uparrow}^+ f_{i\downarrow}^+|0,\Psi_{M-1}^{N-1}\rangle
\end{gather}
describes a state with a single particle occupying the distinguished site and its partner residing deeper inside the superconductor. Note that this state characterizes one broken Cooper pair. Due to the coupling to the quantum dots, states with additional broken Cooper pairs could also occur. Since breaking further pairs always costs an energy proportional to $g$, we neglect states with more than one broken pair.

On the subspace spanned by $|\Psi_0\rangle$, $|\Psi_1\rangle$ and $|\Psi_2\rangle$, the Hamiltonian of the superconductor in isolation is now represented by
\begin{gather}
\Hv_{SC+kin} = \left[\begin{array}{ccc} g\nu & \sqrt{2\nu}\,t & \beta \\ \sqrt{2\nu}\,t & g & \sqrt{2(1-\nu)}\,t \\ \beta & \sqrt{2(1-\nu)}\,t & g(1-\nu) \end{array}\right],
\label{eq:hkin}
\end{gather}
where $\beta = -g\sqrt{\nu(1-\nu)}$ as it was defined in Eq. \eqref{eq:beta}. It is instructive to study the probability of single occupancy of the distinguished level as a function of $t$. If $t=0$, the eigenvalues are $0$, $g$ and $g$; the lowest energy of  0 corresponds to a mixture of $|\Psi_0\rangle$ and $|\Psi_2\rangle$ indicating that the probability of single occupancy is zero. If the hopping parameter has a small but non-zero value, the $|\Psi_1\rangle$ state is admixed to the ground state, indicating that the ground state contains broken Cooper pairs. Second order perturbation theory indicates that the ground state energy is shifted to $-2t^2/g$ in the presence of a small $t$. In the opposite limit, $t\rightarrow \infty$, the eigenvalues are $-\sqrt{2}t$, $0$ and $\sqrt{2}t$. The ground state is $\big(|\Psi_0\rangle -\sqrt{2}|\Psi_1\rangle + |\Psi_2\rangle\big)/2$. In this state, the probability of having a single particle on the distinguished level is $1/2$.
Between the two limiting cases, the probability increases smoothly from $0$ to $1/2$.

It should be emphasized that in reality singly occupied levels do not appear in the GS of superconductors in isolation. Blocked levels in the Richardson model are only present in the excited states \cite{vonDelft2001} or due to the Cooper-pair breaking action of magnetic impurities. The results are thus physically meaningful only when the QD is coupled to the superconductor and the local magnetic moment breaks pairs even in the GS.

Taking also the dots into account, one has to add further states with one broken Cooper pair to all sectors investigated in the flatband limit. The calculation of the Hamiltonian on these extended Hilbert spaces is tedious but straightforward and can be found in the Appendix~\ref{appD}.

\begin{figure}[ht]
\centering
\includegraphics[width=9cm]{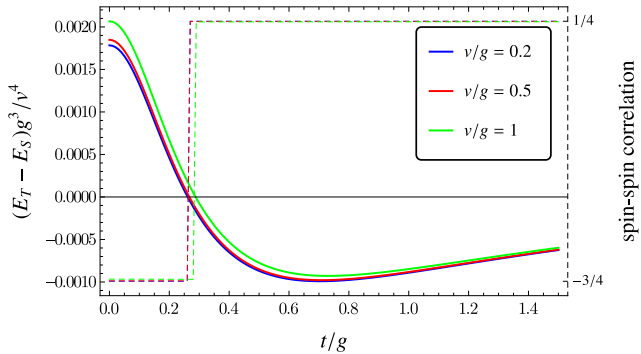}
\caption{Comparison of minimal energies in the singlet ($E_S$) and the triplet ($E_T$) sector at different values of hybridization strength, analytical results. Solid lines illustrate the energy difference while dashed lines show the spin-spin correlation between the dots in the ground state. By exceeding some limit value of the bandwidth, the ground state changes from singlet to triplet. The transition value is around $0.3g$ and  depends only slightly on the hybridization strength.}
\label{fig:dessminmod}
\end{figure}

The minimal energies of the obtained Hamiltonians are computed numerically and the difference between the minimal energies are shown in Fig. \ref{fig:dessminmod}. The results show that for small values of $t$, the ground state forms a singlet, while by increasing the strength of the kinetic energy, the ground state becomes a triplet. This is further confirmed by calculating the spin-spin correlation between the two quantum dots which suddenly switches from approximately -3/4 to close to 1/4 as the levels of singlet and triplet states cross each other.

The minimal model for even number of particles implies that if the hopping processes between superconducting energy levels are large enough, i.e., the kinetic energy term is significant, then it is energetically  favourable to break a Cooper pair inside the superconductor and allow the dot spins to form a triplet through the mediation of one member of the broken pair. This is also the ground state for a normal-state bath in the double quantum dot impurity problem \cite{vojta2002,hofstetter2002,zitko2006,pustilnik2006,zitko2007,logan2009,zitko2012}.

\section{Charging energy}

In the case of a small SI with a large charging energy $E_C$, we add $E_C(\hat{N}_{SC} - n_0)^2$ to the Hamiltonian, where $\hat{N}_{SC}$ is the SI occupancy operator and $n_0$ is the target number of particles controlled in experimental setups by tuning the gate voltage applied on the SI.

\begin{figure}[ht]
\centering
\includegraphics[width=8cm]{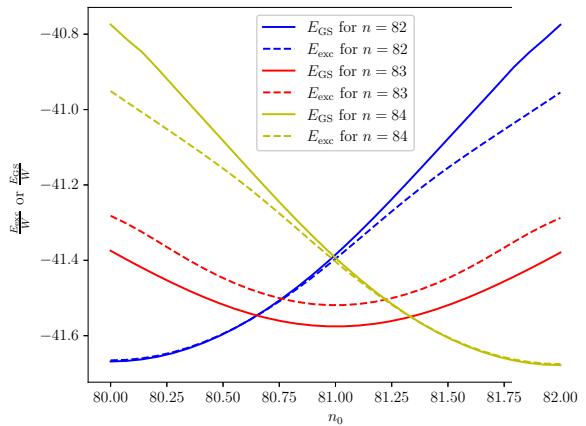}
\caption{Energy of the ground and first excited state as a function of $n_0$ for states with a different total number of particles $n$. For $n=83$, the ground state is the parity-antisymmetric state with FM alignment, and the excited state is symmetric and AFM aligned. For even number of particles, solid lines represent the triplet configuration with FM alignment, dashed lines correspond to singlet states.}
\label{fig:n0dep}
\end{figure}

We study by using DMRG whether the spin configuration of the dots changes when tuning $n_0$. The parameters were chosen to fit typical experimental setups, $U=6\Delta$, $\varepsilon=-3\Delta$, $E_C=2\Delta$. We use $g=0.8W$, which corresponds to $\Delta=0.16W$. For the hybridization, $v=0.25W$ is set which is equivalent to $\Gamma/U=0.102$ with $\Gamma=\pi v^2/(2W)$ the hybridization strength. The number of sites is $N=80$ and the the number of total particles are $n=82$, 83 and 84. Fig.~\ref{fig:n0dep} shows the $n_0$-dependence of the energy of both the ground and the first excited state.For odd number of particles, the solid line indicates the parity anti-symmetric sector with FM alignment while the excited state is the symmetric, AFM aligned state. It is readily seen that the charging energy only shifts these energy states.

For even number of particles, however, one can observe level-crossing. The ground state turns from an FM aligned triplet state to an AFM singlet when $n_0$  differs from $n$ the actual number of particles. The kink in the solid lines occurs where the energy of the subgap state reaches the gap edge.

The results indicate that the charging energy can also have an effect on the spin configuration. The detailed behavior depends both on the ratios of various energy scales $E_C$, $U$, $g$, $W$, and also on the gate voltage through $n_0$. Exploring these dependencies is beyond the scope of the present paper.

\section{Conclusion}

In this paper, the system consisting of a superconducting island and two quantum dots is modelled within the framework of the charge-conserving Richardson model. We have focused on the regime when both dots are singly occupied with a spin-carrying particle and studied how the two spins are effectively coupled to each other. For odd number of particles, the spins align ferromagnetically. We find this result both in the flatband limit and in the full model with finite bandwidth. For even number of particles, the spins align antiferromagnetically in the flatband limit and ferromagnetically in the wideband limit, with a transition between the two cases at some intermediate ratio of bandwidth to pairing strength. 
This significant change in the nature of the ground state implies that the flatband limit results are not fully generic and cannot be extended to any finite bandwidth case, not even at qualitative level. This is in contrast with what has been found for a single QD case \cite{rok2022}.

On the technical side, we introduced an extension of the DMRG method for solving the general problem of two QDs coupled through a SI for arbitrary parameters.
We furthermore presented a minimal analytical model that provides qualitatively adequate description of the spin configurations even for a finite bandwidth. 

Finally, we note that our results are also relevant for applications aiming to split Cooper pairs by coupling a superconductor to two quantum dots. We have shown that the singlet configuration (AF alignment) of the QD spins is the ground state only in a limited part of the parameter space and can be realized only when the pairing interaction is strong. 

\begin{acknowledgments}
We acknowledge the support of the Slovenian Research Agency (ARRS) under P1-0416 and J1-3008.
\end{acknowledgments}


\appendix
\section{L\" owdin orthogonalization of superconductor levels}
\label{appA}

We present a unitary transformation of the SC levels leading to a more convenient form of the hybridization.
In general, the hybridizations are given by
\begin{gather}
H_{hyb,L} = v_L \sum_{i} \left(\gamma_{Li} d_{L\uparrow}^+ c_{i\uparrow} + \gamma_{Li}^* d_{L\downarrow}^+ c_{i\downarrow} + h.c. \right), \nonumber \\
H_{hyb,R} = v_R \sum_{i} \left(\gamma_{Ri} d_{R\uparrow}^+ c_{i\uparrow} + \gamma_{Ri}^* d_{R\downarrow}^+ c_{i\downarrow} + h.c. \right)\,.
\end{gather}
The operators $\sum_{i}\gamma_{Li}c_{i\uparrow}$ and $\sum_{i}\gamma_{Ri}c_{i\uparrow}$ describe one (two) distinguished state(s) if the overlap parameter fulfills $|\alpha|=1$ ($|\alpha|<1$). Here, $\alpha = \sum_i \gamma_{Li}^*\gamma_{Ri}$.

First, we study the case of $|\alpha|<1$ with two distinguished states which are not necessarily orthogonal to each other.
We choose the unitary transformation such that two of the new basis vectors lie in the subspace of the two distinguished states, $\mathcal{D}$, while others form an orthonormal system which is also orthogonal to $\mathcal{D}$. Inside $\mathcal{D}$, we apply L\" owdin orthogonalization \cite{lowdin1,piela_orth} to obtain a symmetric form of the hybridization. 

Let us introduce the operators
\begin{gather}
f_{L\uparrow}' = \sum_{i}\gamma_{Li}c_{i\uparrow}, \nonumber \\
f_{R\uparrow}' = \sum_{i}\gamma_{Ri}c_{i\uparrow}
\end{gather}
which are not orthogonal for $\alpha\neq0$. The L\" owdin orthogonalization is performed as
\begin{gather}
f_{L\uparrow} = r f_{L\uparrow}' - p^* f_{R\uparrow}' \nonumber \\
f_{R\uparrow} = - p f_{L\uparrow}' + r f_{R\uparrow}'
\end{gather}
with
\begin{gather}
r=\frac{\sqrt{1+|\alpha|}+\sqrt{1-|\alpha|}}{2\sqrt{1-|\alpha|^2}}, \nonumber \\
p=e^{i\delta}\frac{\sqrt{1+|\alpha|}-\sqrt{1-|\alpha|}}{2\sqrt{1-|\alpha|^2}}\,.
\end{gather}
The phase factor occurs if the overlap parameter is a complex number, $\alpha = |\alpha|e^{i\delta}$. After the transformation, the hybridizations can be rewritten as
\begin{widetext}
\begin{gather}
H_{hyb,L}= v_L\sqrt{1-|\alpha|^2}\left(rd_{L\uparrow}^+ f_{L\uparrow} + p^*d_{L\uparrow}^+f_{R\uparrow} + rd_{L\downarrow}^+ f_{L\downarrow} + pd_{L\downarrow}^+f_{R\downarrow} + h.c.\right), \nonumber \\
H_{hyb,R}= v_R\sqrt{1-|\alpha|^2}\left(pd_{R\uparrow}^+ f_{L\uparrow} + rd_{R\uparrow}^+f_{R\uparrow} + p^*d_{R\downarrow}^+ f_{L\downarrow} + rd_{R\downarrow}^+f_{R\downarrow} + h.c.\right)\,.
\end{gather}
\end{widetext}
The phase of $\alpha$ can be scaled out with the gauge transformation
\begin{gather}
d_{L\uparrow} = \tilde{d}_{L\uparrow}e^{-i\frac{\delta}{2}} \qquad d_{R\uparrow} = \tilde{d}_{R\uparrow}e^{i\frac{\delta}{2}} \nonumber \\
f_{L\uparrow} = \tilde{f}_{L\uparrow}e^{-i\frac{\delta}{2}} \qquad f_{R\uparrow} = \tilde{f}_{R\uparrow}e^{i\frac{\delta}{2}} \nonumber \\
d_{L\downarrow} = \tilde{d}_{L\downarrow}e^{i\frac{\delta}{2}} \qquad d_{R\downarrow} = \tilde{d}_{R\downarrow}e^{-i\frac{\delta}{2}} \nonumber \\
f_{L\downarrow} = \tilde{f}_{L\downarrow}e^{i\frac{\delta}{2}} \qquad f_{R\downarrow} = \tilde{f}_{R\downarrow}e^{-i\frac{\delta}{2}}\,.
\end{gather}
Henceforth, we will drop the notation of $\sim$ and consider that both $\alpha$ and $p$ are equal to their absolute value.
By introducing the angle $\varphi$ as $\sin(2\varphi) = \alpha$, we obtain
\begin{gather}
H_{hyb,L}= v_L\sum_{\sigma}\left(\cos(\varphi)d_{L\sigma}^+ f_{L\sigma} + \sin(\varphi)d_{L\sigma}^+f_{R\sigma}+ h.c.\right), \nonumber \\
H_{hyb,R}= v_R\sum_{\sigma}\left(\sin(\varphi) d_{R\sigma}^+ f_{L\sigma} + \cos(\varphi)d_{R\sigma}^+f_{R\sigma} + h.c.\right)\,.
\end{gather}
This hybridization is formally the same as that considered in Ref.~\cite{filip}.

The great advantage of this Hamiltonian is that in the special case of $U_L=U_R$, $\varepsilon_L = \varepsilon_R$ and $v_L=v_R$, it has a parity symmetry. The symmetry ensures that the system decouples to a symmetric and an anti-symmetric sectors.

In the case of $\alpha=1$, there is no need for L\" owdin orthogonalization. The unitary transformation must be chosen such that one of the new basis vectors is described by $f_{0\sigma} = \sum_{i} \gamma_{L/Ri}c_{i\sigma}$, while the other new basis vectors must only fulfill orthonormality.

\section{Matrix product operator representation of the Hamiltonian}
\label{MPO}

\begin{widetext}

The matrix product representation of the Hamiltonian is given as $H=\prod_{j=0}^{N+1} W_j$.
The first element, $W_0$, is a vector and defines the site with the left quantum dot:
\begin{equation}
W_0 = \begin{pmatrix}
I &
[ \epsilon \hat{n}_L + U \hat{n}_{L\uparrow} \hat{n}_{L\downarrow} ]&
0 &
0 &
0 &
0 &
d_{L\uparrow}   &
d_{L\downarrow} &
d^\dagger_{L\uparrow}   &
d^\dagger_{L\downarrow} &
0 &
0 &
0 
\end{pmatrix}.
\end{equation}
$d_{L\sigma}$ are the left quantum dot operators, $\hat{n}_{L\sigma} = d^\dagger_{L\sigma} d_{L\sigma}$, and  $\hat{n}_L = \hat{n}_{L\uparrow} + \hat{n}_{L\downarrow}$.
The superconducting sites $j=1,\ldots,N$ are represented by matrices $W_j$:
\begin{equation}
W_j = \begin{pmatrix}
1 & X_j & - v_R \gamma_{Rj} c_{j\uparrow} F_j & - v_R \gamma_{Rj}^* c_{j\downarrow} F_j & v_R \gamma_{Rj}^* c^\dagger_{j\uparrow} F_j & v_R \gamma_{Rj} c^\dagger_{j\downarrow} F_j & 0 & 0 & 0 & 0 & g y_j c_{j\downarrow} c_{j\uparrow} & g y_j c^\dagger_{j\uparrow} c^\dagger_{j\downarrow} & 2 E_c n_j  \\
0 & I & 0 & 0 & 0 & 0 & 0 & 0 & 0 & 0 & 0 & 0 & 0 \\
0 & 0 & F_j & 0 & 0 & 0 & 0 & 0 & 0 & 0 & 0 & 0 & 0 \\
0 & 0 & 0 & F_j & 0 & 0 & 0 & 0 & 0 & 0 & 0 & 0 & 0 \\
0 & 0 & 0 & 0 & F_j & 0 & 0 & 0 & 0 & 0 & 0 & 0 & 0 \\
0 & 0 & 0 & 0 & 0 & F_j & 0 & 0 & 0 & 0 & 0 & 0 & 0 \\
0 & v_L \gamma_{Lj}^* c^\dagger_{j\uparrow} & 0 & 0 & 0 & 0 & F_j & 0 & 0 & 0 & 0 & 0 & 0 \\
0 & v_L \gamma_{Lj} c^\dagger_{j\downarrow} & 0 & 0 & 0 & 0 & 0 & F_j & 0 & 0 & 0 & 0 & 0 \\
0 & v_L \gamma_{Lj} c_{j\uparrow} & 0 & 0 & 0 & 0 & 0 & 0 & F_j & 0 & 0 & 0 & 0 \\
0 & v_L \gamma_{Lj}^* c_{j\downarrow} & 0 & 0 & 0 & 0 & 0 & 0 & 0 & F_j & 0 & 0 & 0 \\
0 & y_j c^\dagger_{j\uparrow} c^\dagger_{j\downarrow} & 0 & 0 & 0 & 0 & 0 & 0 & 0 & 0 & 1 & 0 & 0 \\
0 & y_j c_{j\downarrow} c_{j\uparrow} & 0 & 0 & 0 & 0 & 0 & 0 & 0 & 0 & 0 & 1 & 0 \\
0 & n_j & 0 & 0 & 0 & 0 & 0 & 0 & 0 & 0 & 0 & 0 & 1 
\end{pmatrix},
\end{equation}
where the on-site potential term is
\begin{equation}
    X_j = [\epsilon_j + E_c (1-2 n_0)] \hat{n}_j + (y_j^2 g + 2 E_c) \hat{n}_{j\uparrow} \hat{n}_{j\downarrow}.
\end{equation}
$F_j = (-1)^{\hat{n}_j}$ is the local fermionic parity operator. It takes care of the fermionic anticommutation rules by giving a factor of $-1$ if the $j$-th level is occupied by a single particle.
Finally, the right quantum dot is represented by a vector $W_{N+1}$:
\begin{equation}
W_{N+1} = \begin{pmatrix}
[ \epsilon_R + U n_{R\uparrow} n_{R\downarrow}) n_R ] &
 1 &
 d^\dagger_{R \uparrow}  &
 d^\dagger_{R \downarrow}  &
 d_{R \uparrow}  &
 d_{R \downarrow}&
 0 &
 0 &
 0 &
 0 &
 0 &
 0 &
 0 
\end{pmatrix}.
\end{equation}

\end{widetext}

\section{Superexchange between two dots}
\label{appC}

In this section, we consider a single level mediating exchange interaction between two QDs. 
Each dot hosts a single particle and we assume symmetric configuration of parameters.
The interaction then depends on the occupancy of the intermediate level.

The Hamiltonian is given as
\begin{equation}
\begin{split}
H=&\varepsilon\sum_{\sigma}\left(d^{+}_{L\sigma} d_{L\sigma} + d^{+}_{R\sigma} d_{R\sigma} \right) + U\sum_{I=L,R} n_{I\uparrow}n_{I\downarrow} \\
&+ \varepsilon_{f}\sum_{\sigma}f^{+}_\sigma f_\sigma + v \sum_\sigma\left(d^{+}_{L\sigma} f_\sigma + d^{+}_{R\sigma} f_\sigma  + h.c.\right),
\end{split}
\end{equation}
where the on-site interaction $U$ may be finite. Note that in the pairing model of the main text, there is no on-site potential term; that situation corresponds to $\epsilon_f=0$. 
In what follows, we systematically study all cases depending on the occupancy of the intermediary level and compare the AF and FM configurations. \\

\underline{Zero occupancy:}\\

For FM aligned case, a representative state is given as $|\uparrow,\uparrow\rangle\otimes|0\rangle$, where the first part $|\uparrow,\uparrow\rangle$ describes the state of the two dots, while the second part is the state of the intermediate level. The basis vectors in this spin sector are 
\begin{gather}
|\uparrow,\uparrow\rangle\otimes|0\rangle, \nonumber \\
\frac{1}{\sqrt{2}}\left(|\uparrow,0\rangle - |0,\uparrow\rangle\right) \otimes |\uparrow\rangle \,.
\end{gather}
The Hamiltonian is represented by
\begin{gather}
\mathbf{H}_{T_0} = \left[\begin{array}{cc} 2\varepsilon & \sqrt{2} v \\ \sqrt{2} v & \varepsilon + \varepsilon_f \end{array}\right],
\label{eq:t0supexch}
\end{gather}
where the index $T_0$ indicates that the system is an overall triplet state.

\begin{figure}[ht]
\centering
\includegraphics[width=8cm]{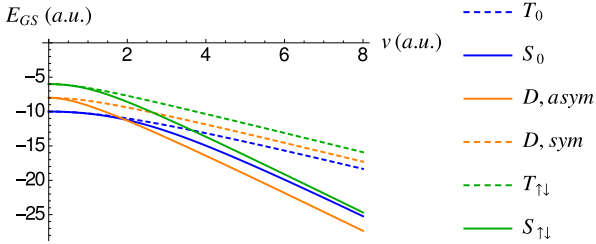}
\caption{Comparison of minimal energies of anti-ferromagnetically (solid line) and ferromagnetically (dashed line) aligned configurations in case of different occupancies of the intermediate level. For the plot, $\varepsilon_f = 2$, $U=10$ and $\varepsilon=-5$ were chosen in arbitrary units.}
\label{fig:supexch}
\end{figure}

For AF aligned case, the basis vectors are
\begin{gather}
\frac{1}{\sqrt{2}}\left(|\uparrow,\downarrow\rangle - |\downarrow,\uparrow\rangle\right) \otimes |0\rangle, \nonumber \\
\frac{|\uparrow,0\rangle + |0,\uparrow\rangle}{2}\otimes |\downarrow\rangle - \frac{|\downarrow,0\rangle + |0,\downarrow\rangle}{2}\otimes |\uparrow\rangle, \nonumber \\
|0,0\rangle\otimes |\ud\rangle, \\
\frac{1}{\sqrt{2}} \left(|\ud,0\rangle + |0,\ud\rangle\right) \otimes |0\rangle,
\label{eq:s0supexch}
\end{gather}
which describe overall singlet states.
On this subspace, the Hamiltonian is represented by
\begin{gather}
\mathbf{H}_{S_0} = \left[\begin{array}{cccc} 2\varepsilon & \sqrt{2} v & 0 & 0 \\ \sqrt{2} v & \varepsilon + \varepsilon_f & 2v & \sqrt{2}v \\ 
0 & 2v & 2\varepsilon_f & 0 \\
0 & \sqrt{2}v & 0 & 2\varepsilon + U \end{array}\right].
\end{gather}
By calculating the minimal energy for both $T_0$ and $S_0$, we find that the singlet sector has lower minimal energy, see the blue lines in Fig. \ref{fig:supexch}.\\

\underline{Single occupancy:}\\

In this case, the $f$ level hosts one particle with spin-down. The total spin is 1/2 and, hence, the sector describes doublet states but we distinguish between parity symmetric and anti-symmetric subsectors exhibiting AF and FM alignment on the dots, respectively.
In the parity-anti-symmetric sector, the basis vectors are given by
\begin{gather}
\frac{1}{\sqrt{2}}\left(|\uparrow,0\rangle - |0,\uparrow\rangle \right)\otimes |\ud\rangle, \nonumber \\
\frac{1}{\sqrt{6}}\left(2|\uparrow,\uparrow\rangle\otimes |\downarrow\rangle - \left(|\uparrow,\downarrow\rangle + |\downarrow,\uparrow\rangle \right)\otimes |\uparrow\rangle\right), \nonumber \\
\frac{1}{\sqrt{2}}\left(|\uparrow,\ud\rangle - |\ud,\uparrow\rangle\right) \otimes |0\rangle, \nonumber \\
\frac{1}{\sqrt{2}}\left(|0,\ud\rangle - |\ud,0\rangle\right) \otimes |\uparrow\rangle,
\end{gather}
and the Hamiltonian is represented by
\begin{gather}
\mathbf{H}_{D,asym} = \left[\begin{array}{cccc} \varepsilon + 2\varepsilon_f & \sqrt{3} v & 0 & v  \\ 
\sqrt{3} v & 2\varepsilon + \varepsilon_f & \sqrt{3} v & 0 \\
0 & \sqrt{3} v & 3\varepsilon + U & v \\
v & 0 & v & 2\varepsilon + U + \varepsilon_f \end{array}\right]\,.
\end{gather}

In the parity symmetric sector, the basis vectors are defined by
\begin{gather}
\frac{1}{\sqrt{2}}\left(|\downarrow,0\rangle + |0,\downarrow\rangle\right) \otimes |\ud\rangle, \nonumber \\
\frac{1}{\sqrt{2}}\left(|\uparrow,\downarrow\rangle - |\downarrow,\uparrow\rangle\right) \otimes |\downarrow\rangle, \nonumber \\
\frac{1}{\sqrt{2}}\left(|\ud,\downarrow\rangle + |\downarrow,\ud\rangle\right)\otimes |0\rangle, \nonumber \\
\frac{1}{\sqrt{2}}\left(|\ud,0\rangle + |0,\ud\rangle\right)\otimes |\downarrow\rangle,
\end{gather}
and the Hamiltonian reads
\begin{gather}
\mathbf{H}_{D,sym} = \left[\begin{array}{cccc} \varepsilon + 2\varepsilon_f & -v & 0 & -v \\ -v & 2\varepsilon + \varepsilon_f & -v & 0 \\
0 & -v & 3\varepsilon + U & v \\ -v & 0 & v & 2\varepsilon + U + \varepsilon_f
 \end{array}\right]\,.
\end{gather}
By calculating the minimal energy for both $D_{sym}$ and $D_{asym}$, we find that the sector $D_{asym}$ featuring ferromagnetic alignment has lower minimal energy, see the red lines in Fig. \ref{fig:supexch}. \\

We note that the quadruplet states (e.g. $S_z=3/2$) are always higher in energy.

\underline{Double occupancy:}\\

This sector is analogous to the case of zero occupancy. The $f$ level now hosts two particles. In the triplet sector, the basis states are
\begin{gather}
|\uparrow,\uparrow\rangle \otimes |\ud\rangle, \nonumber \\
\frac{1}{\sqrt{2}}\left( |\ud,\uparrow\rangle - |\uparrow,\ud\rangle\right)\otimes |\uparrow\rangle,
\end{gather}
with the Hamiltonian
\begin{gather}
\mathbf{H}_{T_{\ud}} = \left[\begin{array}{cc} 2\varepsilon + 2\varepsilon_f & \sqrt{2} v \\ \sqrt{2} v & 3\varepsilon + U + \varepsilon_f \end{array}\right]\,.
\end{gather}

In the singlet sector, we have the following states.
\begin{gather}
\frac{1}{\sqrt{2}}\left(|\uparrow,\downarrow\rangle - |\downarrow,\uparrow\rangle\right) \otimes |\ud\rangle, \nonumber \\
\frac{1}{2}\left( (|\ud,\downarrow\rangle + |\downarrow,\ud\rangle)\otimes |\uparrow\rangle - (|\ud,\uparrow\rangle + |\uparrow,\ud\rangle)\otimes |\downarrow\rangle \right), \nonumber \\
\frac{1}{\sqrt{2}} \left(|\ud,0\rangle + |0,\ud\rangle\right)\otimes |\ud\rangle, \nonumber \\
|\ud,\ud\rangle \otimes |0\rangle.
\end{gather}
The Hamiltonian is represented by the matrix
\begin{gather}
\mathbf{H}_{S_{\ud}} = \left[\begin{array}{cccc} 2\varepsilon + 2\varepsilon_f & \sqrt{2}v & 0 & 0 \\ \sqrt{2} v & 3\varepsilon + U + \varepsilon_f & -\sqrt{2} v & -2 v \\ 0 & -\sqrt{2} v & 2\varepsilon + U + 2\varepsilon_f & 0 \\ 0 & -2 v & 0 & 4\varepsilon + 2U \end{array}\right]\,.
\end{gather}
As in the zero occupancy sector, the singlet has lower minimal energy, see the green lines in Fig. \ref{fig:supexch}.

\begin{figure}[ht]
\centering
\includegraphics[width=6.5cm]{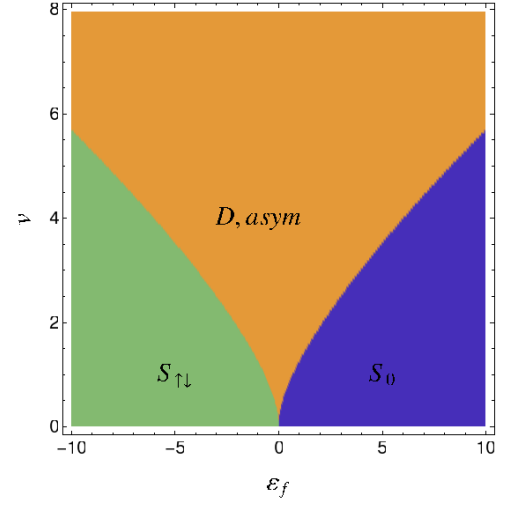}
\caption{Phase diagram of the ground state at $\varepsilon = -5$ and $U=10$.}
\label{fig:mapsl}
\end{figure}

So far, we have seen that for zero/single/double occupancy of the intermediating level, the sector $S_0$/$D_{asym}$/$S_{\ud}$ has the lowest minimal energy. The phase diagram in Fig. \ref{fig:mapsl} shows the sector with the lowest overall energy as a function of $v$ and $\varepsilon_f$.

\section{Minimal analytical model of kinetic energy to capture}
\label{appD}
The kinetic energy can be taken into account with the hopping process between the distinguished level and all other levels.
The corresponding Hamiltonian is defined as
\begin{gather}
H_{kin} = \frac{t}{\sqrt{N}} \sum_{j\neq 0,\sigma} \left(f_{0\sigma}^+ f_{j\sigma} + h.c.\right),
\end{gather}
which is added to the Hamiltonian in the flat-band limit as given in Eq. \eqref{eq:hamf}.
We note that, in general, one could consider hopping terms between all remaining levels but that would break analytical solvability. We shall see that this minimal model is sufficient to capture the singlet-triplet transition as a function of bandwidth.

In this Appendix, we study only the case of an even number of particles because this is the only sector where the DMRG shows level-crossing.

First, we consider the states in the singlet sector. The set of basis states of Eq. \eqref{eq:singletstates} must be extended as follows:
\begin{widetext}
\begin{equation}
\begin{split}
|\phi_{0}^S\rangle & = \frac{1}{\sqrt{2}} \left(|\uparrow,\downarrow\rangle - |\downarrow,\uparrow\rangle\right) \otimes |0,\Psi^{N-1}_{M}\rangle , \\
|\phi_{1}^S\rangle & = \frac{1}{2}\left(|\uparrow,\downarrow\rangle - |\downarrow,\uparrow\rangle\right) \otimes \left( |\uparrow,\Psi^{N-1}_{M-1+\downarrow}\rangle - |\downarrow,\Psi^{N-1}_{M-1+\uparrow}\rangle\right)  , \\
|\phi_{2}^S\rangle & = \frac{1}{\sqrt{2}} \left(|\uparrow,\downarrow\rangle - |\downarrow,\uparrow\rangle\right) \otimes |\ud,\Psi_{M-1}^{N-1}\rangle , \\
|\phi_{-1X}^S\rangle & = \frac{1}{2}\Big( \left( |\uparrow,0\rangle + |0,\uparrow\rangle \right) \otimes |0,\Psi^{N-1}_{M+\downarrow}\rangle - \left(|\downarrow,0\rangle + |0,\downarrow\rangle\right) \otimes |0,\Psi^{N-1}_{M+\uparrow}\rangle\Big) , \\
|\phi_{0X}^S\rangle & = \frac{1}{2}\Big((|\uparrow,0\rangle + |0,\uparrow\rangle) \otimes |\downarrow,\Psi_{M}^{N-1}\rangle - (|0,\downarrow\rangle + |\downarrow,0\rangle) \otimes |\uparrow,\Psi_{M}^{N-1}\rangle \Big) , \\
|\phi_{1X}^S\rangle & = \frac{1}{2}\Big( \left( |\uparrow,0\rangle + |0,\uparrow\rangle \right) \otimes |\ud,\Psi^{N-1}_{M-1+\downarrow}\rangle - \left(|\downarrow,0\rangle + |0,\downarrow\rangle\right) \otimes |\ud,\Psi^{N-1}_{M-1+\uparrow}\rangle\Big) , \\
|\phi_{-2D}^S\rangle & = |0,0\rangle\otimes |0,\Psi_{M+1}^{N-1}\rangle , \\
|\phi_{-1D}^S\rangle & = \frac{1}{\sqrt{2}}|0,0\rangle\otimes \left( |\uparrow,\Psi_{M+\downarrow}^{N-1}\rangle - |\downarrow,\Psi_{M+\uparrow}^{N-1}\rangle\right) , \\
|\phi_{0D}^S\rangle & = |0,0\rangle \otimes |\ud,\Psi_{M}^{N-1}\rangle.
\end{split}
\end{equation}
On this space, the Hamiltonian is represented by the block matrix
\renewcommand{\arraystretch}{2}
\begin{gather}
\Hv = E_{0}\Iv + \left[\begin{array}{c|c|c}
2\varepsilon + \Hv_{SC+kin} & \begin{array}{ccc} 0 & \sqrt{2}v & 0 \\ 0 & 0 & -v \\ 0 & 0 & 0 \end{array} & 0 \\
\hline
\begin{array}{ccc} 0 & 0 & 0 \\ \sqrt{2}v & 0 & 0 \\ 0 & -v & 0 \end{array} & \varepsilon + \tilde{\Hv}_{SC+kin} & \begin{array}{ccc} 0 & \sqrt{2} v & 0 \\ 0 & 0 & 2v \\ 0 & 0 & 0 \end{array} \\
\hline
0 & \begin{array}{ccc} 0 & 0 & 0 \\ \sqrt{2} v & 0 & 0 \\ 0 & 2v & 0 \end{array} & g(2\nu-1) + \Hv_{SC+kin}
\end{array}\right],
\end{gather}
\renewcommand{\arraystretch}{1}
where $\Hv_{SC+kin}$ is defined in Eq. \eqref{eq:hkin} and
\begin{gather}
\tilde{\Hv}_{SC+kin} = \left[\begin{array}{ccc}
2g\nu & t\sqrt{1-\nu} & \beta\\
 t\sqrt{1-\nu} & g\nu & -t\sqrt{\nu} \\
 \beta & -t\sqrt{\nu} & g
\end{array}\right]\,.
\end{gather}
We note that in principle the operator $H_{kin}$ leads out of this space and we should also take into account states in which two Cooper pairs are broken. However, as we mentioned before, these are very high in energy, therefore we neglect them.

In the triplet sector, the basis states of Eq.~\eqref{eq:tripletstates} must be extended as follows.
\begin{equation}
\begin{split}
|\phi_0^T\rangle & = |\uparrow,\uparrow\rangle \otimes |0,\Psi_{M}^{N-1}\rangle,  \\
|\phi_1^T\rangle & = \frac{1}{\sqrt{2}}|\uparrow,\uparrow\rangle \otimes\Big( |\uparrow,\Psi_{M-1+\downarrow}^{N-1}\rangle - |\downarrow,\Psi_{M-1+\uparrow}^{N-1}\Big) , \\
|\phi_2^T\rangle & = |\uparrow,\uparrow\rangle \otimes |\ud,\Psi_{M-1}^{N-1}\rangle , \\
|\phi_1'^T\rangle & = \frac{1}{2}\left[|\uparrow,\uparrow\rangle \otimes\Big( |\uparrow,\Psi_{M-1+\downarrow}^{N-1}\rangle + |\downarrow,\Psi_{M-1+\uparrow}^{N-1}\Big) - \Big(|\uparrow,\downarrow\rangle + |\downarrow,\uparrow\rangle\Big) \otimes |\uparrow,\Psi_{M-1+\uparrow}^{N-1}\rangle\right] , \\
|\phi_{-1X}^T \rangle & = \frac{1}{\sqrt{2}}\Big(|\uparrow,0\rangle - |0,\uparrow\rangle\Big) \otimes |0,\Psi_{M+\uparrow}^{N-1}\rangle , \\
|\phi_{0X}^T \rangle & = \frac{1}{\sqrt{2}}\Big(|\uparrow,0\rangle - |0,\uparrow\rangle\Big) \otimes |\uparrow,\Psi_{M}^{N-1}\rangle , \\
|\phi_{1X}^T \rangle & = \frac{1}{\sqrt{2}}\Big(|\uparrow,0\rangle - |0,\uparrow\rangle\Big) \otimes |\ud,\Psi_{M-1+\uparrow}^{N-1}\rangle.
\end{split}
\end{equation}
We note that these states all belong to the total $S_z = 1$. The sectors corresponding to $S_z=0$ and $S_z =-1$ would lead to the same Hamiltonian matrix.
On this space, the Hamiltonian is represented by
\renewcommand{\arraystretch}{2}
\begin{gather}
\Hv = E_{0}\Iv + \left[\begin{array}{c|c|c}
2\varepsilon + \Hv_{SC} & \begin{array}{c} 0 \\ 0 \\ 0 \end{array} & \begin{array}{ccc} 0 & \sqrt{2} v & 0 \\ 0 & 0 & -v \\ 0 & 0 & 0 \end{array} \\
\hline
\begin{array}{ccc} 0 & 0 & 0 \end{array} & 2\varepsilon + g & \begin{array}{ccc} 0 & 0 & \sqrt{2}v \end{array} \\
\hline
\begin{array}{ccc} 0 & 0 & 0 \\ \sqrt{2}v & 0 & 0 \\ 0 & -v & 0 \end{array} & \begin{array}{c} 0 \\ 0 \\ \sqrt{2}v \end{array} & \varepsilon + \tilde{\Hv}_{SC}
\end{array}\right].
\end{gather}
\renewcommand{\arraystretch}{1}
To compare the singlet and triplet sectors, we obtain the lowest minimal energies within the perturbation theory. For $t,v\ll g\ll\varepsilon$, the energy difference is found to be
\begin{equation}
E_T-E_S = \frac{2v^4}{|\varepsilon|^3}\left(1- \frac{3|\varepsilon|t^2}{2g^3}\right),
\label{eq:etmes}
\end{equation}
indicating that the energy difference decreases with increasing $t$. The decreasing trend is induced by the sixth order process which admixes the state $|\phi_{1}'^T \rangle$, which has a relatively low energy of $2\varepsilon+g$, to the ground state. The sixth order process involves breaking a Cooper pair and a superexchange procedure between the states $|\phi_{1}'^T \rangle$ and $|\phi_{1}^T \rangle$ through one member of the broken pair. This process only exists in the triplet sector since there is no singlet counterpart to the $\phi_{1}'^T$ state. As $t$ increases, the sixth order process competes with the fourth order processes favouring AF alignment and will eventually overtake it 
at $t=t_0=\sqrt{2g^3/3|\epsilon|}$.

Finally, we provide the perturbation theory results for both $E_S$ and $E_T$. We list only the most important sixth order terms, which involve states with both dots occupied since these are the most relevant for large $|\varepsilon|$. For small $v$ and $t$,
\begin{equation}
\begin{split}
E_S &= E_0 + 2\varepsilon + \frac{v^2}{\varepsilon-\frac{g}{2}} - \frac{2t^2}{g} +
\frac{2t^2v^2}{g\left(\varepsilon-\frac{g}{2}\right)^2} - \frac{v^4}{g\left(\varepsilon-\frac{g}{2}\right)^2} + \frac{4t^4}{g^3}+ \frac{v^4}{\varepsilon\left(\varepsilon-\frac{g}{2}\right)^2}
+ E_S^{(6)}\\
E_T &= E_0 + 2\varepsilon + \frac{v^2}{\varepsilon-\frac{g}{2}} - \frac{2t^2}{g} + 
\frac{2t^2v^2}{g\left(\varepsilon-\frac{g}{2}\right)^2} - \frac{v^4}{g\left(\varepsilon-\frac{g}{2}\right)^2} + \frac{4t^4}{g^3} - \frac{v^4}{\left(\varepsilon-\frac{g}{2}\right)^3}
+ E_T^{(6)}  \\
E_S^{(6)} & = \frac{t^2v^4}{g^2(2\varepsilon-g)}\left( \frac{1}{\left(\varepsilon-\frac{g}{2}\right)^2} + \frac{1}{\left(\varepsilon-\frac{3g}{2}\right)^2} - \frac{1}{\left(\varepsilon-\frac{g}{2}\right)\left(\varepsilon-\frac{3g}{2}\right)}\right) \\
E_T^{(6)} & = \frac{t^2v^4}{g^2(-g)}\left( \frac{1}{\left(\varepsilon-\frac{g}{2}\right)^2} + \frac{1}{\left(\varepsilon-\frac{3g}{2}\right)^2} + \frac{1}{\left(\varepsilon-\frac{g}{2}\right)\left(\varepsilon-\frac{3g}{2}\right)}\right)
\end{split}
\end{equation}
and in the limit of $g\ll |\varepsilon|$, we obtain
\begin{equation}
\begin{split}
E_S &= E_0 + 2\varepsilon + \frac{v^2}{\varepsilon} - \frac{2t^2}{g} + \frac{2t^2v^2}{g^2\varepsilon} - \frac{v^4}{g\varepsilon^2}+ \frac{4t^4}{g^3} + \frac{v^4}{\varepsilon^3} + \frac{t^2v^4}{2g^2 \varepsilon^3}\\
E_T &= E_0 + 2\varepsilon + \frac{v^2}{\varepsilon} - \frac{2t^2}{g} + \frac{2t^2v^2}{g^2 \varepsilon} - \frac{v^4}{g\varepsilon^2}+ \frac{4t^4}{g^3} - \frac{v^4}{\varepsilon^3} - \frac{3t^2v^4}{g^3\varepsilon^2}
\end{split}
\end{equation}
which is reformulated in Eq.~\eqref{eq:etmes} in the limiting case of $g\ll|\varepsilon|$.

\end{widetext}

\bibliographystyle{apsrev}
\bibliography{flatband}

\end{document}